\documentclass[doublecol]{epl2} 
\usepackage{graphicx}
\usepackage[activeacute,english]{babel}
\begin{document}

\title{Two-state flashing molecular pump}

\author{A. Gomez-Marin\footnote{Current address: EMBL-CRG Systems Biology Unit, Center for Genomic Regulation, UPF - Barcelona, Spain, EU.} and J. M. Sancho}
\institute{Facultat de F\'isica, Universitat de Barcelona - Diagonal
647, Barcelona, Spain, EU}

\abstract{Here we study a pumping device capable of maintaining a density gradient and a flux of particles 
across a membrane. Its driving mechanism is based on the flashing ratchet effect powered by  the random telegraph process in the presence of thermal fluctuations. Unlike Brownian motors, 
the concentrations at both reservoir boundaries
need to be implemented as boundary conditions.
The residence transition rates of the dichotomic flashing are related to the binding and hydrolysis of ATP molecules. 
The model is exactly solved and explored. The pump energetics is discussed and the relevant parameter values are tuned within a biological scale.
}


\maketitle

{\bf Introduction.}
Active transport processes are ubiquitous in living matter. In fact, the cell cannot survive when specific out-of-equilibrium tasks inside cease, mainly carried out by proteins such as kinesin, RNA-isomerase, dynein or rotatory ATP-ase.
For instance, in order to maintain a suitable osmotically regulated state, the pumping of ions across the cell membrane 
is a crucial job that is performed by molecular pumps, which are, in turn, powered by the hydrolysis of ATP molecules.
Stemming from the studies on active transport in noisy environments \cite{rei}, the stochastic features of pumps were originally addressed in \cite{prost,ast0} and applied to chemically driven electron pumping \cite{ast}.
An experiment on artificial ion pores showing rectification was presented in \cite{siw1}, where the observations where interpreted by a model containing flashing ratchets.
Also in  \cite{siw2}, experimental realizations of synthetic nanoscopic devices transporting potassium ions against their concentration gradient when stimulated with external field fluctuations were built.
Recently it has been directly demonstrated that the ratchet mechanism is indeed of use in a biological system where molecular transport driven by fluctuations occurs in a membrane channel \cite{kos}.
Other theoretical works have studied the 
gating kinetics of ionic channels, based on the effect of non-equilibrium fluctuations combined with the ratchet effect \cite{lee}. 
A Brownian device with cyclic steps accounting for its configurational changes has been shown
to reproduce the main features of experiments on the Na,K-ATP-ase ion pump \cite{chang}.
Moving to more simplistic descriptions, the study of active transport in an idealized Brownian pump focusing on the concentration gradient created and maintained at both sides of a membrane has been 
analytically and numerically discussed in \cite{agm}. 

Despite its relevance for modeling molecular pumps,
the ratchet effect and its transport features against a concentration gradient have drawn very little attention.
In this Letter we introduce a simple, yet revealing, two-state theoretical model for a 
machine which pumps particles against a concentration gradient. 
The model is inspired by recent experimental works where the operating cycle of a channel has been shown to involve two steps (opening and closing) related to the ATP binding and hydrolysis respectively \cite{berger, gadsby}.
The pumping device we present consists of a ratchet potential, embedded in a membrane and bounded by particle reservoirs, which exhibits dichotomic fluctuations controlled by the ATP concentration.
We do not focus on the precise structural details of the pump, but do it at the level of energetics. 
We consider the asymmetric ratchet profile as an energy barrier.
In the steady state, we exactly calculate the concentration ratio between both reservoirs and the particle flux. We relate the performance of the device with ATP concentration, study its energetics and finally place it in a biological context.

\begin{figure}
\begin{center}
  \includegraphics[angle=0, width=6cm]{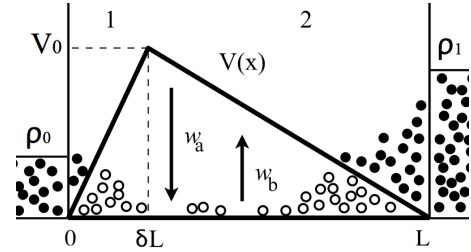}
  \caption{Pump model: an asymmetric potential $V(x)$ 
 flashes its shape at random between a blocking (closed) and a flat (open) configuration  
with transitions rates $w_a$ and $w_b$.}
\label{model}
\end{center}
\end{figure}

{\bf The model.}
Let us consider a pumping mechanism based on a flashing  potential embedded in a membrane of length $L$, whose boundaries are infinite reservoirs of particles with fixed mean densities $\rho_0$ and $\rho_1$. The dynamics of a particle in the membrane is determined by the following Langevin equation in the over-damped regime,
\begin{equation} \label{langpump2}
\gamma \dot{x}=- U'(x,t)+\xi(t),
\end{equation}
where $U(x,t)$ is a time dependent potential (the prime denotes position derivative) and $\xi(t)$ the random force arising from the 
thermal fluctuations. It has zero mean and its autocorrelation is given by the fluctuation-dissipation theorem,
\begin{equation}
\langle \xi(t)\xi(t')\rangle=2\gamma k_{B}T \delta(t-t'),
\end{equation}
where $\gamma$ is the friction coefficient and $T$ the temperature of the environment.
Whereas in the reservoirs the particles can diffuse freely, in the membrane they experience the potential
\begin{equation}
U(x,t)=V(x)\zeta(t),
\end{equation}
where  $V(x)$, as depicted in figure \ref{model}, is a piecewise linear asymmetric ratchet potential:
\begin{eqnarray}
V_{1}(x)=V_{0} \frac{x}{\delta L}, &  & x \; \in \; [0,\delta L],\\
V_{2}(x)=V_{0} \frac{L-x} {(1-\delta)L}, &   & x\; \in \; [\delta L, L].
\label{potential}
\end{eqnarray}
$V_0$ is the energy barrier height and $\delta$ controls the spatial asymmetry by taking values between zero and unity.

The flashing modulation $\zeta(t)$ is a stochastic forcing in time which randomly switches the ratchet
potential between two states \cite{rei,prost}.
It is distributed according to a random telegraph (dichotomous Markov) process, flipping between values $\zeta_{a}=1$ (closed state) and $\zeta_{b}=0$ (open state), with residence probability
transitions $w_{a}$ and $w_{b}$ respectively \cite{shapiro,gard}. Its mean value and correlation are
\begin{eqnarray}
\langle \zeta(t) \rangle & = & \frac{w_{a}}{w_{a}+w_{b}},\\
\langle  \Delta \zeta(t) \Delta \zeta(t')  \rangle & = & \frac{w_{a}w_{b}}{(w_{a}+w_{b})^{2}}e^{-(w_a +w_b)|t-t'|}
\end{eqnarray}
where $\Delta \zeta(t) \equiv \zeta(t)-\langle \zeta(t)\rangle$.
While the asymmetric potential breaks the spatial symmetry, the flashing transitions between states are responsible for the breaking of detailed balance in the system. 

\begin{figure} [t]
\begin{center}
  \includegraphics[angle=0, width=6cm]{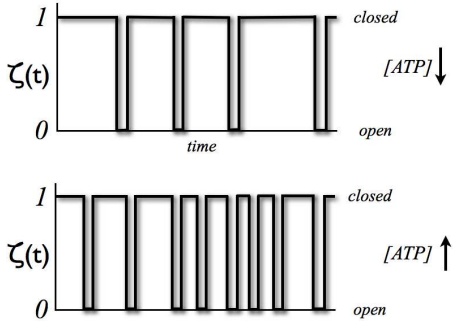}
      \caption{Illustrative representation of  the effect of ATP concentration in the switching
rates of the dichotomic flashing ratchet mechanism. }
    \label{ATPdixh} 
\end{center}
\end{figure}

{\bf Transition rates and ATP concentration.}
The hydrolysis of ATP molecules provides the energy input. It creates the necessary out-of-equilibrium conditions to drive molecular pumps.
As a first level of approximation to include the effect of ATP consumption into the model, 
we relate the ATP concentration, its binding and its hydrolysis processes to both residence transition rates.
This two-state picture has been described in experiments in channels \cite{berger,gadsby} and it is a valid paradigm for the mathematical characterization of mechano-chemical energy transduction.
We consider that, when the pump is in the closed state ($\zeta_a=1$), it is the binding  of an ATP molecule to a certain pocket center of the pump  which induces conformational change to the open state ($\zeta_b=0$). 
The transition rate $w_a$ is thus dependent on the ATP concentration available in the surroundings, which can be modeled as a Michaelis-Menten kinetics law:
\begin{equation}
w_{a}=w_0 \frac{[\rm{ATP}]}{k_M+[\rm{ATP}]} =  \frac{w_0}{1+1/\sigma},
\label{M-M}
\end{equation}
where $k_M$ is the so-called affinity constant and $[\rm{ATP}]$ is the concentration of ATP molecules. Only their ratio is relevant: $\sigma \equiv [\rm{ATP}]/k_M$.
The rate $w_b$ is given by the inverse of the machine's intrinsic time, which we consider constant:
\begin{equation}
w_b=w_0.
\end{equation}
In the limit of a saturating $[\rm{ATP}]$, both transition probabilities are equal ($w_a=w_b=w_0$) indicating that the pump cannot operate faster.  
Medium $[\rm{ATP}]$ values imply a longer average time in the closed state. 
At low concentrations, $w_{a}$ is small and the pump seldom starts the cycle. Diffusive leak losses dominate and its pumping capacity is then very low.  
See scheme in figure \ref{ATPdixh}.


\begin{figure}
\begin{center}
    \includegraphics[angle=270, width=6.6cm]{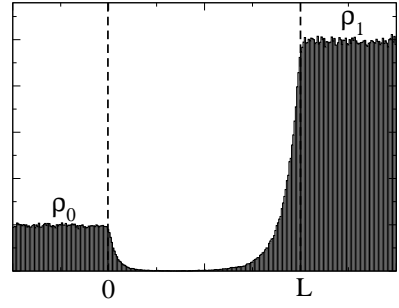}
  \caption{
  Density profile obtained directly from numerical simulations of the Langevin equation in the steady state at zero flux.}
\label{fig1pump2}
\end{center}
\end{figure}

{\bf Numerical simulations.} 
As a first test to show that the model just introduced can indeed create and maintain a concentration gradient, we use the simulation framework discussed in \cite{agm}.
This method allows for the numerical implementation of the reservoir boundary conditions, which is not trivial.
We then simulate non-interacting particles following the Langevin equation (\ref{langpump2}), under thermal white noise and the random telegraph process.
In figure \ref{fig1pump2}, we plot 
the histogram of the position of the particles, being illustrative of the density profiles typically  obtained in the steady state at zero net flux.
It is clear that the pump is able to generate and maintain a concentration ratio ($\rho_1 \neq  \rho_0 $), while the edges emulate particle reservoirs with uniform concentrations.

{\bf Theoretical analysis.} 
Consistent with the Langevin equation (\ref{langpump2}), 
the corresponding partial differential equations for the time evolution of the
concentration of particles in each flashing state, $P_a \equiv P_a(x,t)$ and $P_b \equiv P_b(x,t)$, 
originate from a combination of the Fokker-Planck description for the continuous variable $x$ and the Master equation for the discrete states $a$ and $b$. They read
\cite{prost,shapiro,gard} 
\begin{eqnarray}
\partial_{t}P_{a}  & = &  \gamma^{-1}  \partial_{x} [V'(x)+k_{B}T\partial_{x}]P_{a}-w_{a}P_{a}+w_{b}P_{b}, \nonumber \\
\partial_{t}P_{b}  & = & \gamma^{-1} \partial_{x} [k_{B}T\partial_{x}]P_{b}-w_{b}P_{b}+w_{a}P_{a}.
\label{maineq}
\end{eqnarray}
It is convenient to work with dimensionless quantities.
We first re-scale space and time: $z\equiv x/L$ and $s\equiv t (k_B T)/(\gamma L^2)$. Consequently, the new transition rates are $f \equiv w (\gamma L^2)/(k_B T)$ (for $w_a$, $w_b$ and $w_0$) and $L$ is absorbed in the concentration distributions.
We introduce the following densities: the total density of particles
$\rho \equiv P_{a}+P_{b}$  and the auxiliary function $\pi \equiv P_{a}-P_{b}$.
In the steady state, eqs. (\ref{maineq}) are governed by a new set of equations for each region (indicated by the subindex $i=1,2$):
\begin{eqnarray} \label{P}
\rho_i^{'}  -F_i  [\rho_i+\pi_i]  & = & -J,\\
\label{Q} 
\pi_i^{''}-F_i (\rho_i^{'}+\pi_i^{'}) &= &\pi(f_{b}+f_{a})-  \rho(f_{b}-f_{a}),
\end{eqnarray}
where $ F_1= - v_0/2\delta$ and $F_2=v_0/2(1-\delta)$.  
$v_0\equiv V_0/k_BT$ measures the relative strength of the potential with respect to the thermal energy.
$J$ is the physical total flux of particles across the membrane.

Substituting (\ref{P}) in (\ref{Q}) one gets a linear ODE for $\rho_{i}(z)$,
\begin{equation} \label{main}
 \rho^{'''}_i(z)-2F_i \rho^{''}_i(z) -(f_{a}+f_{b})\rho^{'}_i(z)+2f_{b}F_i
\rho_i(z) =J(f_a+f_b),
\end{equation}
whose solution has the standard form
\begin{eqnarray} \label{formRho}
\rho_{i}(z) = 
C_{i1}e^{\lambda_{i1}z}+C_{i2}e^{\lambda_{i2}z}+C_{i3}e^{\lambda_{i3}z} +(J/F_i)\alpha, 
\end{eqnarray}
where $\alpha \equiv (f_a+f_b)/2f_b$. 
The constants $C_{ij}$ are, so far, unknown parameters. 
Instead, the constant coefficients $\lambda_{ij}$ are found 
by inserting $\rho_{i}(z)$ back in (\ref{main}) and solving  the algebraic equation
 \begin{equation}
 \lambda_{ij}^{3} - 2F_{i}\lambda_{ij}^{2} - (f_{a}+f_{b})\lambda_{ij} +
 2f_{b}F_{i} =0.  
\label{lambdas}
\end{equation}
Plugging (\ref{formRho}) in equation (\ref{P}) and defining 
$s_{ij} \equiv (\lambda_{ij}/F_i -1)$ and $\beta \equiv (f_b-f_a)/2f_b$,
one finds
\begin{equation} \label{formQ}
\pi_{i}(z) = s_{i1} C_{i1}e^{\lambda_{i1}z} +s_{i2} C_{i2}e^{\lambda_{i2}z} +
s_{i3}  C_{i3}e^{\lambda_{i3}z}+(J/F_i)\beta.
\end{equation}

{\bf Boundary conditions.} The formal solution of the concentration of particles in the steady state 
depends on seven unknown constants (the coefficients $C_{ij}$ and the flux $J$), whose value is determined by boundary conditions.
In contrast with models for Brownian motors, in the present Brownian pump one should not impose normalization nor the periodicity condition for the probability.
Instead, the total concentrations at left and right boundaries, $\rho_0$ and $\rho_1$ respectively, are externally fixed:
$\rho_1 (0)\equiv  \rho_0$ and $\rho_2(1)\equiv  \rho_1$.
At the boundaries (in contact with the particle reservoirs) the state probabilities can be factorized as the total concentration times the stationary weight the state, complying with $P_a(0)=\rho_0 f_b/(f_b+f_a)$ and 
$P_b(0)=\rho_0 f_a/(f_b+f_a)$ (and analogously at $z=1$).
Rewritten in terms of $\pi(z)$, this supplies two new conditions:
$\pi_1(0)=r \rho_0$ and $\pi_2(1)= r \rho_1$,
where $r\equiv(f_b-f_a)/(f_b+f_a)$.
Third, $\rho(z)$ and $\pi(z)$ must be continuous functions connecting both regions:
$\rho_{1}(\delta)=\rho_{2}(\delta)$ and $\pi_{1}(\delta)=\pi_{2}(\delta)$.
Last, the total flux $J$ associated to $\rho(z)$ has implicitly been assumed continuous across both zones. The same holds for the one associated to $\pi(z)$ in (\ref{Q}), which in practice yields to  
impose continuity on $\pi_i'(z)-F_i[\rho_i(z)+\pi_i(z)]$.
This last boundary condition can be recast as a property of the $\pi(z)$ derivatives: $(F_{2}-F_{1})[\rho_{1}(\delta)+\pi_{1}(\delta)]
=\pi_{2}'(\delta)-\pi_{1}'(\delta).$

{\bf Solution.}
The above cumbersome boundary conditions can be compactly written as a system of linear equations: 
\begin{eqnarray}
\rm{see\; eq. (\ref{bigmat}),} \nonumber
\end{eqnarray}
\begin{widetext}
\begin{equation} \label{bigmat}
\left(
\begin{array}{ccccccc}
1 & 1 & 1 & 0 & 0 & 0 & \alpha /F_1 \\

0 & 0 & 0 & e^{-\lambda_{21}} & e^{-\lambda_{22}} & e^{-\lambda_{23}} & \alpha /F_2 \\

s_{11} & s_{12}  & s_{13} & 0 & 0 & 0 & \beta /F_1 \\

0 & 0 & 0 & s_{21} e^{\lambda_{21}} & s_{22} e^{\lambda_{22}} &  s_{23} e^{\lambda_{23}}    & \beta / F_2\\

e^{\lambda_{11}\delta} & e^{\lambda_{12}\delta} & e^{\lambda_{13}\delta}& -e^{\lambda_{21}\delta}& -e^{\lambda_{22}\delta}&-e^{\lambda_{23}\delta} & \alpha \Delta \mathcal{F}\\

s_{11}  e^{\lambda_{11}\delta} &  s_{12}  e^{\lambda_{12}\delta} & s_{13} e^{\lambda_{13}\delta} & - s_{21}  e^{\lambda_{21}\delta} & - s_{22}  e^{\lambda_{22}\delta}
& - s_{23}  e^{\lambda_{23}\delta} & \beta \Delta \mathcal{F}\\

h_{11} e^{\lambda_{11}\delta} &  h_{12} e^{\lambda_{12}\delta} & h_{13} e^{\lambda_{13}\delta} & 
-  \lambda_{21} s_{21} e^{\lambda_{21}\delta} & - \lambda_{22} s_{22} e^{\lambda_{22}\delta} &
-  \lambda_{23} s_{23} e^{\lambda_{23}\delta} & F_2 \Delta \mathcal{F} 
\end{array}
\right) 
\left(
\begin{array}{c}
C_{11}\\
C_{12}\\
C_{13}\\
C_{21}\\
C_{22}\\
C_{23}\\
 J
\end{array}
\right)
=
\left(
\begin{array}{c}
\rho_0\\
\rho_1\\
r \rho_0\\
r \rho_1\\
0\\
0\\
0
\end{array}
\right)
\end{equation}
\begin{equation}\label{bigdet}
\rm{det} 
\left |
\begin{array}{cccccc}
 \rho_1/\rho_0 & \rho_1/\rho_0 & \rho_1/\rho_0 & e^{-\lambda_{21}} & e^{-\lambda_{22}} & e^{-\lambda_{23}} \\

s_{11}-r & s_{12}-r  & s_{13}-r & 0 & 0 & 0 \\

0 & 0 & 0 & (s_{21}-r) e^{\lambda_{21}} & (s_{22}-r) e^{\lambda_{22}} &  (s_{23}-r) e^{\lambda_{23}} \\

e^{\lambda_{11}\delta} & e^{\lambda_{12}\delta} & e^{\lambda_{13}\delta}& -e^{\lambda_{21}\delta}& -e^{\lambda_{22}\delta}&-e^{\lambda_{23}\delta} \\

s_{11}  e^{\lambda_{11}\delta} &  s_{12}  e^{\lambda_{12}\delta} & s_{13} e^{\lambda_{13}\delta} & - s_{21}  e^{\lambda_{21}\delta} & - s_{22}  e^{\lambda_{22}\delta}
& - s_{23}  e^{\lambda_{23}\delta} \\

h_{11} e^{\lambda_{11}\delta} &  h_{12} e^{\lambda_{12}\delta} & h_{13} e^{\lambda_{13}\delta} & 
-  \lambda_{21} s_{21} e^{\lambda_{21}\delta} & - \lambda_{22} s_{22} e^{\lambda_{22}\delta} &
-  \lambda_{23} s_{23} e^{\lambda_{23}\delta} 

\end{array}
\right |
=0
\end{equation}
\end{widetext}
from which the flux $J$ (and also the density profile $\rho(z)$ by means of $C_{ij}$) can be obtained as a function of the system parameters.
New constants are defined for compact notation: $\Delta \mathcal{F} \equiv (1/F_1-1/F_2 )$ and $h_{ij} \equiv \lambda_{ij} \left(  s_{ij}+F_{2}/F_{1}-1 \right)$.
In the zero flux case ($J=0$) such linear system has to be rearranged.
The ratio of concentrations $\rho_1/\rho_0$, which is then the relevant observable, is easily found by imposing 
a minor of the coefficient matrix equal to zero (ensuring that the solution of the problem exists):
\begin{eqnarray}
\rm{see\; eq. (\ref{bigdet}).} \nonumber
\end{eqnarray}


{\bf Parameter exploration: zero flux case.}
When the pump is forced to operate at vanishing particle flux,
$\rho_1$ and $\rho_0$ become boundary conditions, whose ratio is the relevant quantity to explore as a function of the parameters of the system.
When examining the behavior of $\rho_{1}/\rho_{0}$ as a function of 
the asymmetry parameter $\delta$, we check that for $\delta=0.5$  (spatially symmetric system) no net pumping is possible, thus leading to $\rho_1 = \rho_0$.  
Moreover, the solution of the problem is anti-symmetric when $\delta$ and $1-\delta$ are interchanged. 
The pumping ability clearly depends on the relative energetic barrier $v_0$.
For low values of $v_0$, the ratchet can barely pump the particles because diffusive losses dominate, whereas as the barrier height is increased, the concentration ratio does so.
For very high values of $v_0$, the fluctuating potential does not improve its effect anymore and its pumping capacity saturates. 
All these features were verified (data not shown) and used as consistency tests for the analytical solution.

\begin{figure} [t]
\begin{center}
  \includegraphics[angle=270, width=6.2cm]{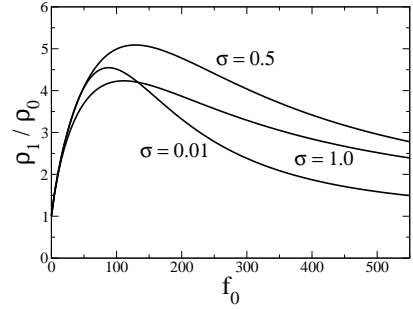}
     \caption{Concentration ratio $\rho_1/\rho_0$ versus the characteristic 
frequency $f_0$ for different $\sigma$ values ($\delta=0.33$ and $v_0=10$). }
  \label{vm}
\end{center}
\end{figure}

The ratio $\rho_1/\rho_0$ as a function of the transition rates of the dichotomic noise is explored in figure \ref{vm}.
By means of varying $f_0$ we observe that there is an optimum value which maximizes the particle gradient.
For a very fast switching ($f_0$ high), the particles do not have time to diffuse in the open configuration and cannot be biased in the closed configuration. 
The opposite limit is also physically intuitive. 
In the adiabatic case ($f_0$ small), no particle gradient can be created: the flashing time-scale is much greater than diffusion, which prevents any possible separation of particles induced by the ratchet effect. 
Further adjustments on $\delta$, $v_0$ and $\sigma$ may allow to obtain even higher concentration ratios, optimizing the performance of the pump.
Figure \ref{vm} depicts how tuning both parameters $f_a$ and $f_b$ (equivalently we vary $f_0$ and $\sigma$) can increase the ratio $\rho_1/\rho_0$.
For very small $[\rm{ATP}]$, $f_a$ tends to zero. The ratchet is in the closed state and it takes a very long to open. This leads to $\rho_1 \to \rho_0$.
For high values of $\sigma$, the rates are equal, $f_a = f_b=f_0$, and $f_0$ alone controls the pumping capacity.
In the intermediate region, there is a maximum distinctive of an optimal behavior, since $\sigma$ can independently improve the pumping capacity given a fixed value of $f_0$.

\begin{figure} 
\begin{center}
    \includegraphics[angle=270, width=6.2cm]{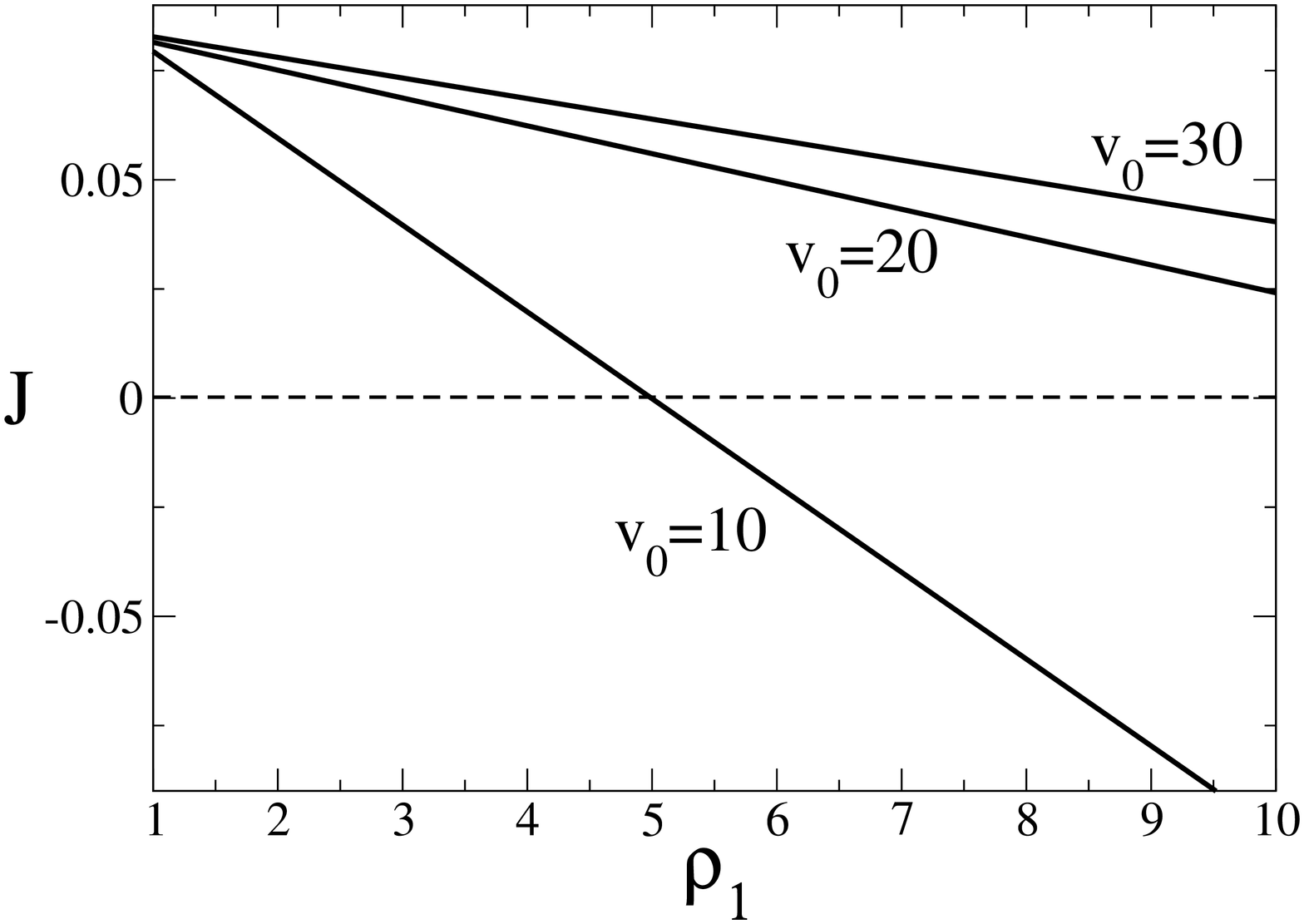}
   \includegraphics[angle=270, width=6.2cm]{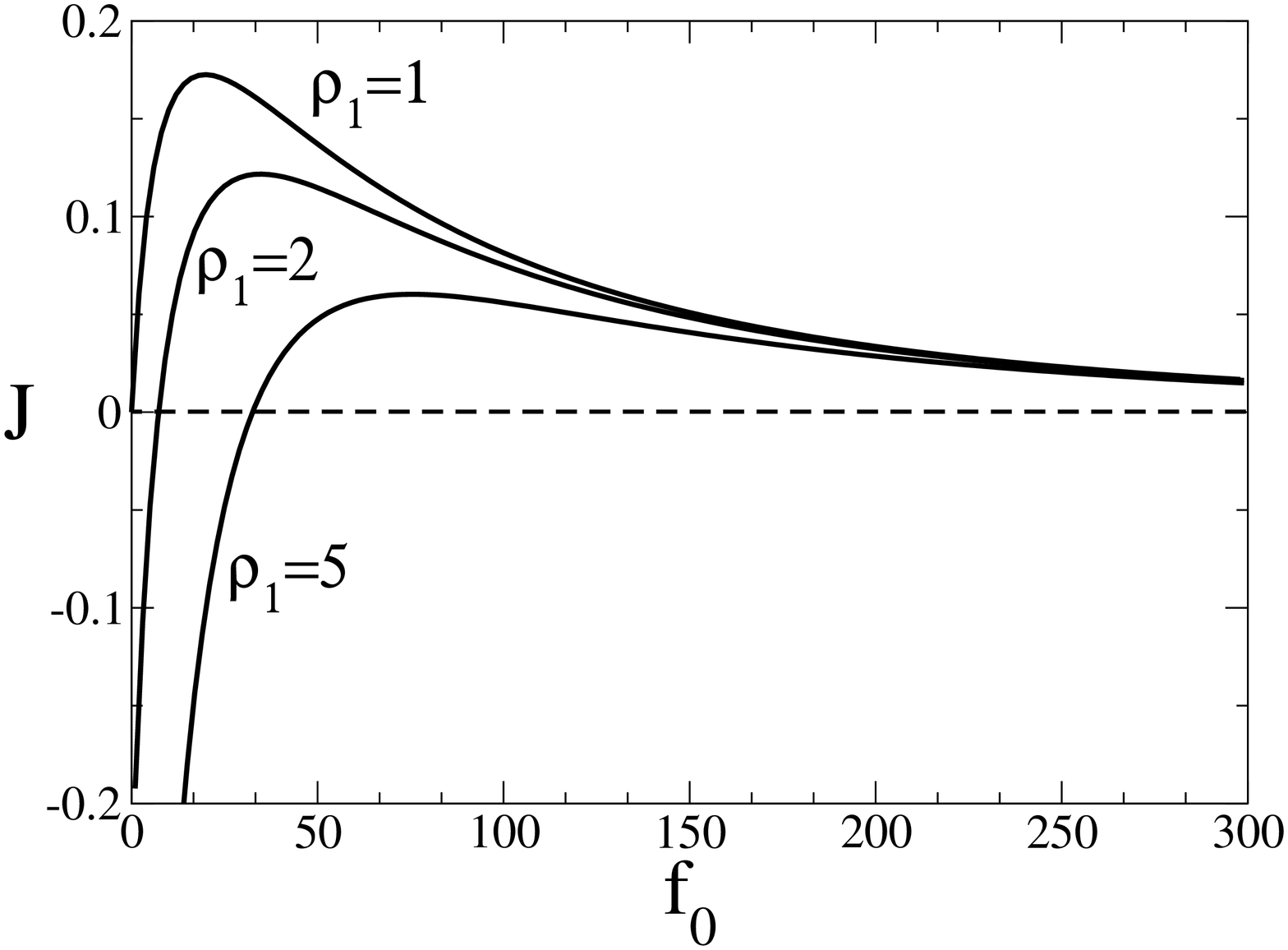}
  \caption{Top: Total flux $J$ as a function of concentration $\rho_1$ when the effective height of the barrier $v_0$ is varied ($\rho_0=1$, $\delta=0.33$, $f_0=100$ and $\sigma=0.5$).
  Bottom:  Flux $J$ as a function of the flashing rate $f_0$ at different concentrations $\rho_1$ ($\rho_0=1$, $\delta=0.33$, $v_0=20$ and $\sigma=0.5$).}
  \label{J1}
\end{center}
\end{figure}

{\bf Parameter exploration: non-zero flux case.} 
In figure \ref{J1}-top we plot the flux $J$ as the concentration $\rho_1$ is increased (given a fixed $\rho_0=1$) for several values of $v_0$.
A restoring entropic force appears opposing the pumping and leading to a decrease of the total flux. The decrease of the flux is linear.
When the barrier $v_0$ is higher, the decay is softer due to a more powerful functioning of the pump (at the expense of more energy).
Even if the pump is able to carry a non-zero net flux of particles against the concentration gradient, if $\rho_1$ is sufficiently increased, the particle flux can be reversed.

The analytical prediction of the flux $J$ as a function of the rate $f_0$ for several values of $\rho_1$ (where $\rho_0=1$) is shown in figure \ref{J1}-bottom.
When $\rho_1=\rho_0$ and $f_0=0$, the device is stopped and the net flux vanishes.
As the pumping rate is increased, transport starts. The higher $\rho_1$, the lower the flux $J$, even becoming negative for great concentration values.
In that case, if the flashing rate is increased, the machine can reverse the sign of the flux back to $J>0$ so that 
particles are then pumped against the concentrations imposed by the reservoirs.
Finally, for very high rates the flux steadily tends to zero because, when the flashing is almost instantaneous, there is no effective pumping.

{\bf Energetic characterization.}
The energetic aspects are relevant and abundant in the literature of motor modeling  \cite{parme,parrondo,sch} and should definitely be addressed in their pump counterparts.
A suitable definition of efficiency for pumps is not obvious.
It is unclear how to interpret and quantify the payoff achieved from the pumping process as a whole.
However, from a purely energetic point of view, meaningful quantities can still be defined consistently.

\begin{figure}
\begin{center}
  \includegraphics[angle=270, width=6.5cm]{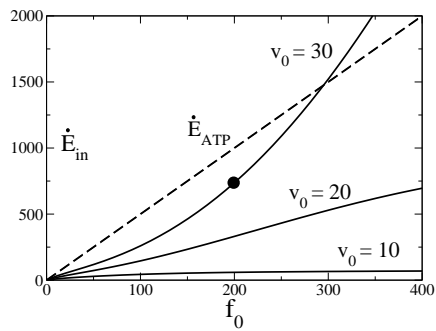} 
  \caption{Input power $\dot{E}_{\rm{in}}$ as a function of 
the flashing rate $f_0$ for three different values of $v_0$ ($\delta=0.33$, $\sigma=0.5$, $\rho_0=1$ and $J=0$). The dashed line represents the power related to the hydrolysis of one ATP molecule per cycle. The parameter space of the model is tuned at the black dot according to biophysical constraints.}
 \label{Einput}
\end{center}
\end{figure}

First, the input source of energy can be conveniently studied as follows.
Working in the overdamped approximation, every time the potential is lifted (due to the $\zeta(t)$ modulation) energy is injected into the system. This energy is used to actively create the concentration profile and maintain it.  When the pump flashes down to the open configuration, no energy is injected nor recovered.
In such situation, the pumping energy input per unit of time is mathematically quantified as 
\begin{equation}\label{Ein}
\dot{E}_{\rm{in}}=\int_{0}^{L}dx V(x) w_b P_b (x).
\end{equation}
The mean power input that the dichotomic flashing inserts into the system is a controllable feature of the model by a judicous choice of its pameter values.
An explicit expression for the above quantity (in dimensionless variables) can be obtained by splitting it into two parts, each corresponding to the piecewise linear potential:
\begin{equation}\label{Ein2}
\frac{\dot{E}_{\rm{in}}}{v_0 f_0}=  \int_{0}^{ \delta}dz  \frac{[\rho_1(z)-\pi_1(z)]/2}{\delta / z} +    \int_{\delta}^{1}dz  \frac{[\rho_2(z)-\pi_2(z)]/2}{(1-\delta)/(1-z)} .
\end{equation}
By using the state probabilities (\ref{formRho}) and (\ref{formQ}) 
we are left with sums of exponentials, whose integrals are trivial.
In figure \ref{Einput} we show the behavior of $\dot{E}_{\rm{in}}$ at zero flux as the typical frequency $f_0$ is varied (solid line) for several values of $v_0$. The faster the flashing, the more energy per unit of time is inserted in the system. Similarly, when the barrier $v_0$ is increased, the power input rises very quickly.

Secondly,  the power output produced by the machine can be quantified as the product ${\dot E}_{\rm{chem}} = J \, \Delta \phi$, where $J$ is the particle current and $\Delta  \phi =  k_B T \ln (\rho_1 / \rho_0)$ is  the chemical potential difference between both reservoirs. 
Note that at zero flux, the power output vanishes, namely, the pump is using energy but it is not performing work. This is equivalent to the stalling regime in motors, where the energetic efficiency is zero since there is consumption but no work performance. 

On the whole, the benefit pursued by a pump is often not an energetic one. For instance, maintaining a concentration gradient at zero flux can indeed be very important for biological functions. The energetic input can still be used to constrain the biophysical tuning of the model, which we address as follows.
In connection to real biomachines, one ought to require the input energy per cycle not to be greater than the energy released from the hydrolysis of one ATP molecule, which is approximately $20k_BT$.
Given that the typical period of the pump cycle is  $\tau_{\rm{cycle}}=w_a^{-1}+w_b^{-1}$,
the power released by the ATP is $ \dot{E}_{\rm{ATP}} = E_{\rm{ATP}}/\tau_{\rm{cycle}}$ which, in dimensionless units, yields
\begin{equation} 
 \dot{E}_{\rm{ATP}} = f_0 \frac{20}{2+1/\sigma} \geq \dot{E}_{\rm{in}}.
\end{equation}
In figure \ref{Einput} we plot the above expression (dashed line) 
as a function of the frequency $f_0$.

{\bf Biophysical scale.}
So far we have solved the equations of the model and examined the behavior of the pump as a function of dimensionless parameters.
Can all that be placed in a biophysical scale \cite{nel,song,kuyu,siw}? 
We use a tuning approach that combines constraints and optimization criteria.
Let us consider the length of the molecular pump of the order of $L \simeq 5nm$ and an effective opening section of $S \simeq 1nm^2$. 
At biological conditions ($T=310K$) the thermal energy is $k_BT \simeq 4.28\cdot 10^{-21}J$.
Given an effective viscosity inside the membrane two orders of magnitude bigger than 
that of the water, the resulting diffusion constant of the medium yields $D=k_BT/\gamma \simeq 10^{-12}m^{2}/s$. 
This leads to a typical time-scale associated with diffusion of the order $\tau =L^2/D \simeq 10^{-5}s$. 
Then, when the optimal fluctuating rate $f_0$ found in our explorations
is mapped to dimensional units we find $w_0 \simeq 10^6 \sim10^7 Hz$,
which is compatible with typical reported rates of functioning of the Na,K-ATP-ase pump under the effect of an oscillating electric field \cite{Hz}.
Guided by the energetic calculations developed previously, we impose that the energy spent per cycle is close but not higher than the one obtained from the ATP hydrolysis. 
See the black dot in figure (\ref{Einput}). 
This sets the possible space of parameters.
For instance, a plausible potential barrier $V_0$ is expected to be of the order of 
a few tens of $k_BT$. Higher values can lead to greater concentration ratios but 
their energetics is not biologically meaningful. 
In this way, by recursively adjusting the parameter space where the pump functions optimally, 
bearing in mind biological constrains,
the model is shown to be consistently translated into dimensional units of the order of those of a real molecular pumps.

{\bf Final remarks and conclusions.} 
The interplay between experimental observations and theoretical models substantially helped to understand the basic features of molecular motors. 
We believe that applying now the basic mechanisms of the ratchet effect to molecular pumps  will contribute to understand the fundamental working principles of protein machines embedded in cell membranes. 
Although the model presented here is a dramatic simplification of the great complexity and richness of a biological pump, it contains many interesting and new elements that capture its basic biophysical aspects.
First, the explicit implementation of concentrations as boundary conditions at both ends of the membrane,  leading to the characteristic  quantity $\rho_1/\rho_0$.
Second, the use of the simulation framework developed in \cite{agm}, which is prone to be applied in models whose an analytical solution is not possible.
Third, a simple coarse-grained connection between transition rates and ATP binding and hydrolysis processes.
Fourth, the energetic characterization of the model and its connection to biophysical values.
Finally, it is worth commenting on the straightforward extension of this approach for particles of charge  $q$ in a membrane potential $\Phi$ by including an electrostatic force $F_q=q\Phi /L$ in the equations of motion. The pump then needs to supply extra energy, reducing the flux and the concentration ratio accordingly.

On the whole, we expect that the present work will be of use as a starting point 
to make progress, from a physical point of view,
towards future investigations of active processes in membranes.

{\bf Acknowledgments.}
We acknowledge financial support from the Ministerio de Educaci\'on y Ciencia of Spain under Project FIS2006-11452-C03-01 and Grant FPU-AP-2004-0770 (A. G--M.).

\end{document}